\documentclass[twocolumn,showpacs,preprintnumbers,amsmath,amssymb,prl]{revtex4}

\usepackage{epsfig}

\newtheorem{lem}{Observation}
\newtheorem{theo}{Theorem}

\begin{document}

\title{Entanglement as precondition for secure quantum key
distribution}
\author{Marcos Curty$^1$, Maciej Lewenstein$^2$, and Norbert L\"{u}tkenhaus$^1$}
\affiliation{$^1$Quantum Information Theory Group, Institut f\"ur
Theoretische Physik, Universit\"{a}t Erlangen-N\"{u}rnberg, 91058
Erlangen, Germany \\
$^2$Institut f\"{u}r Theoretische Physik,
Universit\"{a}t Hannover, 30167 Hannover, Germany}

\begin{abstract}
We demonstrate that a necessary precondition for unconditionally secure
quantum key distribution is that sender and receiver can use the
available measurement results to prove the presence of entanglement in a
quantum state that is
effectively distributed between them. One can thus systematically
search for entanglement using the class of entanglement witness operators
that can be constructed from the observed data. We apply such analysis to
two well-known quantum key distribution protocols, namely the 4-state
protocol and  the 6-state protocol. As a
special case, we show that, for some asymmetric error patterns,
the presence of entanglement can be
proven even for error rates above $25\%$ ($4$-state protocol) and
$33\%$ ($6$-state protocol).
\end{abstract}

\maketitle

Quantum key distribution (QKD) \cite{Wiesner83,BB84} allows two
parties (Alice and Bob) to generate a secret key despite the
computational and technological power of an eavesdropper (Eve), who
interfers with the signals. Together with the Vernam cipher
\cite{Vernam26}, QKD can be used for unconditionally secure
communication.

QKD protocols distinguish typically two phases to establish a key.
In the first phase, an effective  bi-partite quantum mechanical state is
distributed  between the legitimate users, which establishes correlations
between them.  A (restricted) set of measurements is used to measure
these  correlations, and the measurement results are
  described by a joint probability distribution
$P(A,B)$.  In the  second phase, called {\em key distillation}, Alice and
Bob use an authenticated public channel to process the correlated data in
order to obtain a secret key. This procedure involves, typically,
postselection of data, error correction to reconcile the data, and
privacy amplification to decouple the data from a possible eavesdropper
\cite{Norbert99}.

In this Letter we consider  the first phase of QKD and demonstrate
that a necessary precondition for  successful key distillation   is that
Alice  and Bob can
 detect the presence of entanglement in a
quantum state that is effectively distributed between them. Such
detection may involve available observed data only; it can be
realized by using the class of entanglement witness operators that
can be constructed from  these data.

Two types of schemes are typically used to create correlated data.
 In {\em prepare\&measure schemes} (P\&M schemes) Alice
prepares a random sequence of pre-defined non-orthogonal states
that are sent to Bob through an untrusted channel (controlled by
Eve). Generalizing the ideas of Bennett {\it et al.}
\cite{Mermin92}, the signal preparation can be thought of as
follows: Alice prepares an entangled bi-partite state of the form
$|\Psi_{source}\rangle_{AB} =\sum_i \sqrt{p_i}
|e_i\rangle|\varphi_i\rangle$. If she measures the first system in
the canonical orthonormal basis $|e_i\rangle$, she effectively
prepares the (non-orthogonal) signal states $|\varphi_i\rangle$
with probabilites $p_i$. The action of the quantum channel on the
state $|\Psi_{source}\rangle_{AB}$  leads to an effective
bi-partite state shared by Alice and Bob. One important
characteristic of the P\&M  schemes is that the reduced density
matrix $\rho_A$ of Alice is fixed \cite{Note}. In {\em
entanglement based schemes} a bi-partite state is distributed to
Alice and Bob by an, in general,  untrusted third party. This
party may be an eavesdropper who is in possession of a third
sub-system that may be entangled with those given to Alice and
Bob. While the subsystems measured by Alice and Bob result in
correlations described by $P(A,B)$, Eve can use her subsystem to
obtain information about the data of the legitimate users.
Entanglement based schemes have been introduced by Ekert
\cite{Ekert91}, who proposed to detect the presence of
correlations of quantum mechanical nature by looking at possible
violations of Bell-like inequalities. This is, in general,  more
restrictive than detection of the presence of entanglement. As we
will show below, the success of the key distillation phase
requires that the performed measurements together with $P(A,B)$
suffice to prove that the (effective) bi-partite state is
entangled.

The starting point for our considerations is an upper bound for
the distillation rate of a secure key from the correlated data via
public communication, which is given by  the {\em intrinsic
information } $I(A;B\downarrow{}E)$, introduced by Maurer
\cite{Maurer99}. Consider all possible tri-partite states that Eve
can establish using her eavesdropping method, and  all
measurements she could perform on her sub-system. This gives rise
to a set of possible extension ${\cal P}$ of the observable
probability distribution $P(A,B)$ to $P(A,B,E)$. Maurer
\cite{Maurer99} defines $I(A;B\downarrow{}E)$ using the mutual
information $I(A;B|E)$ between Alice and Bob given the public
announcement of Eve's data, as described by the conditional
probabilites $P(A,B|E)$. In an adaptation of Maurer's work we
define the intrinsic information as
\begin{equation}
I(A;B\downarrow{}E) = {\rm inf}_{{\cal P}}\;I(A;B|E) \; .
\end{equation}

An important consequence is that whenever the observable data
$P(A,B)$ can be explained as coming from a tripartite state with a
separable reduced density matrix for Alice and Bob, the intrinsic
information vanishes.
\begin{lem}
Assume that the observable joint probability distribution $P(A,B)$
together with the knowledge of the corresponding measurements
can be interpreted as coming from a separable state
$\sigma_{AB}$. Then the intrinsic information vanishes and no
secret key can be distilled via public communication from the
correlated data.
\end{lem}
This is easy to see for entanglement based schemes as we extend a
separable reduced density matrix $\sigma_{AB}= \sum_i q_i
|\varphi_i\rangle_A\langle \varphi_i|\otimes
|\Phi_i\rangle_B\langle\Phi_i|$ to a tripartite pure state of the
form $|\Psi_{sep}\rangle= \sum_i \sqrt{q_i} |\varphi_i\rangle_A
|\Phi_i\rangle_B|e_i\rangle_E$. Here $|e_i\rangle_E$ is a set of
orthonormal vectors spanning a Hilbert space of sufficient
dimension. If Eve measures her sub-system in the corresponding
basis, the conditional probability distribution conditioned on her
measurement result factorizes so that for this measurement
$I(A;B|E)=0$. As a consequence, the intrinisic information
vanishes and no secret key can be distilled.  In the case of P\&M
schemes we need to show additionally that the state
$|\Psi_{sep}\rangle$ can be obtained by Eve by interaction with
Bob's system only. In the Schmidt decomposition
$|\Psi_{source}\rangle$ can be written as $|\Psi_{source}\rangle =
\sum_i c_i |u_i\rangle_A |v_i\rangle_B$. Then the Schmidt
decomposition of $|\Psi_{sep}\rangle$, with respect of system $A$
and the composite system $BE$, is of the form $|\Psi_{sep}\rangle=
\sum_i\ c_i |u_i\rangle_A |\tilde{e}_i\rangle_{BE}$ since $c_i$
and $|u_i\rangle_A$ are fixed by the known reduced density matrix
$\rho_A$ to the corresponding values of $|\Psi_{source}\rangle$.
Then one can find a suitable unitary operator $U_{BE}$ such that
$|\tilde{e}_i\rangle_{BE}=U_{BE}|v_i\rangle_B|0\rangle_E$ where
$|0\rangle_E$ is an initial state of an auxiliary system.

In both types of schemes it is clear that we can obtain a secret
key whenever the distributed (or effectively distributed)
bi-partite states are entangled qubit states {\em and} we are
allowed to perform joint quantum manipulations on these states.
This is true since one can distill maximally entangled states in
this situation \cite{Horodecki97,Deutsch96}. However, up to now it
is not clear whether this is still true if Alice and Bob perform
their respective measurements and can perform only classical
operations on their correlated data. This scenario has been
partially addressed under extra assumptions in
\cite{Gisin99,Bruss03,Acin03}. More recently, Ac\'in et al.
\cite{Acin03b} proved that one can always distill a secret key from
any bi-partite entangled qubit states by adapting the local
measurements to the quantum state and performing subsequently a
classical protocol.

Let us now turn to the investigation of the correlations in
detail. The question whether the effectively distributed
bi-partite state is entangled can be addressed based on the ideas
of entanglement witnesses.  An entanglement witness is an
observable that detects the presence of entanglement (if any) of a
given state $\rho$: A state $\rho$ is entangled iff there exist a
hermitian witness operator $W$ such that ${\rm Tr}(W\rho)<0$,
while we have ${\rm Tr}(W\sigma)\geq{}0$ for all separable states
$\sigma$. These operators, as any bipartite hermitian operator,
can always be decomposed into a {\it pseudo-mixture} of projectors
onto product vectors $W=\sum_i{}c_i{}\
|a_i\rangle\langle{}a_i|\otimes|b_i\rangle\langle{}b_i|$, where
the coefficients $c_i$ are real numbers and fulfill
$\sum_i{}c_i=1$ \cite{Sanpera97,Guehne02,Guehne02b}. Given such a
decomposition of $W$, the expectation value ${\rm Tr}(W\rho)$ can
be obtained directly from the expectation value, $P(a_i,b_i)$.

In our approach we consider the reverse problem: Given a particular
set of local measurements performed by Alice and Bob, and search for
all entanglement witnesses that can be constructed from them.
\begin{theo}
Given a set of local operations with POVM elements $F_a\otimes G_b$
together with the probability distribution of their occurrence,
$P(A,B)$, then the correlations $P(A,B)$ cannot lead to a secret
key via public communication unless one can prove the presence of
entanglement in the (effectively) distributed state via an
entanglement witness $W=\sum_{a,b} c_{a,b} F_a\otimes G_b$ with
$c_{a,b}$ real such that ${\rm Tr} (W \sigma) \geq 0$ for all
separable states $\sigma$ and $\sum_{a,b} c_{a,b} P(a,b) < 0$.
\end{theo}

By observation 1 it is a necessary condition for the success of
the key distillation phase that we can exclude separable quantum
states as the origin of the observed correlations of the first
phase. The observed data  define equivalence classes of reduced
density matrices that are compatible with the data. We need to
distinguish between cases where the determined equivalence class
contains separable states and those that do not. For this we
proceed as follows: Note that the operators of the form
$\sum_{a,b} c_{a,b} F_a \otimes G_b$ form a real vector space
which is a sub-space of the vector space spanned by the hermitian
operator basis of the composite Hilbert space. The separable
density matrices form a compact, convex set in that vector space,
and its projection  into a sub-space is again a compact and convex
set formed by elements that represent the equivalence classes
mentioned before. Each element of this new set can be explained as
being the projection of a separable density matrix, while the
elements of the complement of  the set cannot be explained in this
way and must therefore come from the projection of an entangled
state.  In the subspace, we therefore need to distinguish a
compact and convex set form its complement, which is done, again,
by witness operators. In the sub-space, all witnesses operators
can be realized by definition. The corresponding operators on the
larger vector space are those witness operators that can be
created by real linear combinations of the local measurements.
This proves the theorem.

The question whether certain correlations are of quantum origin
and might lead to a secure key is therefore reduced to a search
over all entanglement witnesses that can be constructed from the
protocol and the collected data.  We will illustrate the
consequences of this view for well-known protocols, namely the
$6$-state protocol \cite{Bruss98} and  the $4$-state protocol
\cite{BB84}. In searching through the entanglement witnesses, note
that some conditions derived from witnesses are redundant in the
sense that all entangled states detected by one witness can be
contained in the set of detected entangled states of another
witness. A witness operator $W$ is called {\em optimal}
\cite{Lewenstein00} if no other entanglement witness exists that
detects all states detected by $W$. The class of optimal
entanglement witnesses for two qubit states, denoted by OEW, are
given by $W=|\phi_e\rangle\langle\phi_e|^{T_P}$
\cite{Lewenstein97}, where $|\phi_e\rangle$ denotes an entangled
state and ${T_P}$ is the partial trace, that is, the transposition
with respect to one subsystem.

For the case of the 6-state protocol, Alice and Bob perform
projection measurements onto the eigenstates of the three Pauli
operators $\sigma_x, \sigma_y,$ and $\sigma_z$ in the entanglement
based scheme where Eve distributes bi-partite qubit states. In the
corresponding P\&M scheme, Alice prepares the eigenstates of those
operators by performing the same measurements on a maximally
entangled qubit state. Therefore, the set of three measurement
bases used in the protocol allows Alice and Bob to construct any
entanglement witness of the form
\begin{equation}\label{general}
W=\sum_{i,j=\{0,x,y,z\}} c_{ij}\ \sigma_i\otimes\sigma_j,
\end{equation}
where $\sigma_0=\openone$ and $c_{ij}$ are real numbers. Note that
the set of operators $\{\sigma_i\otimes\sigma_j\}_{i,j}$
constitutes an operator basis in
$\mathbb{C}^2\otimes{}\mathbb{C}^2$. Therefore Alice and Bob can
evaluate all entanglement witnesses, in particular the class OEW
of optimal witnesses for two qubits, as given above. This means
that in this protocol all entangled states can be detected.
Alternatively to the witness approach, Alice and Bob can employ
quantum state tomography techniques in connection with the
Peres-Horodecki criterion \cite{Peres96,Horodecki96}.

While the analysis of the $6$-state protocol is quite simple, the
$4$-state protocol, however, needs a deeper examination since it
turns out that the optimal witnesses in OEW cannot be evaluated
with the given correlations, as we will see below. In the four
state protocol Alice and Bob perform projection measurements in
two qubit bases, say $x$ and $z$.  In the corresponding P\&M
scheme Alice uses the same set of measurements on a maximally
entangled state.

For the entanglement scheme we obtain the set of entanglement
witnesses that can be evaluated with the resulting correlations as
\begin{equation}\label{4QKD}
W=\sum_{i,j=\{0,x,z\}} c_{ij}\ \sigma_i\otimes\sigma_j \; .
\end{equation}
This class, which we shall denote as EW$_4$, can be characterized with
the following observation.
\begin{lem}
Given an entanglement witness  $W$ we find  $W\in{}EW_4$ iff $W=W^{T}=W^{T_P}$.
\end{lem}
To see this, we start with the general ansatz of Eqn.~(\ref{general})
and impose the conditions $W=W^{T}=W^{T_P}$. This directly constraints
$W$ to the form (\ref{4QKD}) since $\sigma_y$ is the only
skew-symmetric element in the operator basis. The reverse direction is
then trivial.

Note that that the elements of OEW,
$W=|\phi_e\rangle\langle\phi_e|^{T_P}$, are nonpositive, while
$W^{T_P}=|\phi_e\rangle\langle\phi_e|$ is a positive operator for
all entangled states $|\phi_e\rangle$. This means that, in
contrast to the case of the $6$-state protocol, the $4$-state
protocol does not allow to evaluate the optimal witnesses in OEW.
As a result, there can be  entangled states that give rise to
correlations $P(A,B)$ that are not sufficient to prove the
presence of entanglement.

The concept of optimal witnesses can be extended by calling a
witness $W$ {\em optimal in class $C$} iff there is no other
element in $C$ that detects all entangled states detected by $W$.
Our basic goal is now to characterise a family of witness
operators that are optimal in class EW$_4$, such that it is
sufficient to check this family of witnesses to decide whether the
presence of entanglement can be verified from the given data.

For this purpose we present a necessary condition for a bi-partite
state to contain entanglement that can be detected by elements of
EW$_4$.

\begin{lem}
\label{symmetry}
Given $W\in{}EW_4$, a necessary condition to detect entanglement in
state $\rho$ is that the operator
$\Omega=\frac{1}{4}\left(\rho+\rho^{T_A}+\rho^{T_B}+\rho^T\right)$ is
a non-positive operator.
\end{lem}
To see this, let us start by the observation that the symmetries
of the witness operators in EW$_4$ give rise to the identity ${\rm
Tr}\left(W \rho \right) = {\rm Tr}\left(W \Omega\right)$.  Now let
us assume that the operator $\Omega$ is non-negative. Then one can
interpret it as a density matrix.  Since it is invariant under
partial transposition, it must be a separable state. Since $W$ is
a witness operator, we must therefore find ${\rm Tr}\left(W \rho
\right) \geq 0$. As a result, we find that the non-positivity of
$\Omega$ is a necessary condition to detect entanglement of the
state $\rho$ with witnesses in EW$_4$.

\begin{theo}
Consider the family of operators $W=\frac{1}{2}(Q+Q^{T_P})$, where
$Q=|\phi_e\rangle\langle\phi_e|$ and $|\phi_e\rangle$ denotes a real
entangled state. The elements of this family are witness operators
that are optimal in EW$_4$ and detect all the entangled states that
can be detected within EW$_4$.
\end{theo}

Let us start by checking that this family, indeed, can detect all
entanglement that can be detected in EW$_4$. From the observation
\ref{symmetry} we know that we need only to consider bi-partite
states $\rho_n$ such that $\Omega_n =
\frac{1}{4}\left(\rho_n+\rho_n^{T_A}+\rho_n^{T_B}+\rho_n^T\right)$
is non-positive. We can find, therefore, an (entangled) state
$|\phi_n\rangle$ such that
$\langle\phi_n|\Omega_n|\phi_n\rangle<0$. Moreover, since
$\Omega_n=\Omega_n^T$, this operator has a real representation. In
this representation, also the state $|\phi_n\rangle$ has a real
representation \cite{Strang80}. Let us definine the projector
$Q=|\phi_n\rangle\langle\phi_n|$. Then we find
$\langle\phi_n|\Omega_n|\phi_n\rangle= {\rm
Tr}\left(\frac{1}{4}\left(Q+Q^{T_A}+Q^{T_B}+Q^T\right)\rho
\right)$. Therefore, we can define the operator $W=
\frac{1}{4}\left(Q+Q^{T_A}+Q^{T_B}+Q^T\right)$ that can be further
simplified to $W=\frac{1}{4}\left(Q+Q^{T_P}\right)$ thanks to the
real representation of $Q$. This operator is a witness operator,
since ${\rm Tr}\left(W \sigma \right)\geq 0$ for all separable
states $\sigma$, while ${\rm Tr}\left(W \rho_n \right)< 0$ for the
chosen $\rho_n$. Moreover, by construction the familiy of these
witness operators detect all entanglement that can be detected
within EW$_4$.  The prove of optimality is omitted here.
 $\blacksquare$

The set of witness operators $W=\frac{1}{2}(Q+Q^{T_P})$, with
$Q=|\phi_e\rangle\langle\phi_e|$, provides an infinite number of
necessary and sufficient conditions for the presence of
entanglement in the observable correlations $P(A,B)$. Each
condition is characterized by a real entangled state
$|\phi_e\rangle$, and therefore the coefficients of the
pseudo-mixture decomposition of the corresponding witness
operators can be easily parametrized with three real parameters.
>From a practical point of view, this means that Alice and Bob can
easily check this set of conditions numerically.

Let us briefly analyze the implications of our results in the
relationship between the bit error rate $e$ in the protocols and
the presence of correlations of quantum mechanical nature. Here
error rate refers to the sifted key, that is, those events where
signal preparation and detection employ the same polarization
basis. An intercept/resend attack breaks the entanglement and
gives rise to $e\geq 25\%$ ($4$-state protocol) and $e \geq 33\%$
($6$-state protocol), respectively \cite{Ekert94,Bruss98}. This
means that if the error rate is below these values, this already
sufficies to prove that the joint probability distribution
$P(A,B)$ contains quantum mechanical correlations. However, for
some asymmetric error patterns, it is possible to detect the
presence of quantum correlations even for error rates above $25\%$
($33\%$).  Let us illustrated this fact with an example that is
motivated by the propagation of the polarization state of a single
photon in an optical fiber. This channel can be described by a
unitary transformation that changes on a timescale much longer
than the repetition cycle of the signal source, so it can be
thought to be constant over that time. Consider the unitary
transformation $U(\theta)=\cos\theta\openone-i\sin\theta\sigma_y$.
In this scenario, the resulting bit error rate is given by
$e=\sin^2\theta$ and $e=\frac{2}{3}\sin^2\theta$ for the $4$-state
and the $6$-state protocols, respectively. Nevertheless, in both
cases the existence of quantum correlations can be detected for
all angles $\theta$. The case of the $6$-state protocol is clear,
since a unitary transformation preserves the entanglement and all
entanglement can be verified in this protocol. With respect to the
$4$-state protocol, it can also be shown that there is always an
entanglement witness $W\in$ EW$_4$ that detects quantum
correlations in $P(A,B)$.  In particular, if we select
$W_e=\frac{1}{2}(|\phi_e\rangle\langle\phi_e|+
|\phi_e\rangle\langle\phi_e|^{T_P})$, with $|\phi_e\rangle$ the
eigenvector of the operator
$\frac{1}{2}|\psi\rangle\langle\psi|^{T_P}$
($|\psi\rangle=\cos\theta|00\rangle+
\sin\theta|01\rangle-\sin\theta|10\rangle+\cos\theta|11\rangle$)
which corresponds to its negative eigenvalue, then we find in a
suitable representation as a pseudo-mixture for the entanglement
witness that ${\rm Tr} (W_e \rho) = \sum_i{}c_i{}\
P(a_i,b_i)=-\frac{1}{4}$.

To conclude, we have as a necessary condition for QKD that the
legitimate users can prove the presence of entanglement in the
effectively distributed quantum state. In order to construct
practical and efficient new QKD protocols, it is vital to separate
the generation of two-party correlations from the public
discussion protocol which extracts a key from those data. We have
analyzed the $4$-state and $6$-state QKD protocols, and we have
derived necessary and sufficient conditions for the existence of
quantum correlations in both protocols. As a special case, we have
demonstrated that, for some asymmetric error patterns, the
presence of this type of correlations can be detected even for
error rates above $25\%$ and $33\%$, respectively.

The authors wish to thank D. Bru\ss, P. Hyllus, P. van Loock, Ph.
Raynal, A. Sanpera, and especially O. G\"uhne for very useful
discussions. This work was supported by the DFG (Emmy Noether programme,
SFB 407, SPP 1078), and the
network of competence QIP of the state of Bavaria (A8).

\bibliographystyle{apsrev}

\begin{thebibliography}{99}
\bibitem{Wiesner83} S. Wiesner, Sigact News, {\bf 15}, 78 (1983).
\bibitem{BB84} C. H. Bennett and G. Brassard,  Proc. IEEE Int.
Conference on Computers, Systems and Signal Processing, Bangalore, 
India (IEEE Press, New York, 1984), 175. 
\bibitem{Vernam26} G. S. Vernam, Trans. of the AIEE, {\bf 45}, 295 (1926).
\bibitem{Norbert99} N. L\"utkenhaus, Applied Phys. B, {\bf 69}, 395 (1999).
\bibitem{Mermin92} C. H. Bennett, G. Brassard and N. D. Mermin,
  Phys. Rev. Lett., {\bf 68}, 557 (1992).
\bibitem{Note} That means that Alice has access  to an arbitrary
operator basis for POVM elements of the form
$F_k\otimes\openone_B$, and may include the corresponding  probabilities
in the observable correlations $P(A,B)$. 
\bibitem{Ekert91}  A. K. Ekert, Phys. Rev. Lett., {\bf 67}, 661 (1991).
\bibitem{Maurer99} U. Maurer and S. Wolf, IEEE Trans. Inf. Theory,
  {\bf 45}, 499 (1999).
\bibitem{Horodecki97}  M. Horodecki, P. Horodecki and R. Horodecki,
  Phys. Rev. Lett., {\bf 78}, 574 (1997).
\bibitem{Deutsch96} D. Deutsch, A. Ekert, R. Jozsa, C. Macchiavello,
  S. Popescu and A. Sanpera, Phys. Rev. Lett. {\bf 77}, 2818 (1996).
\bibitem{Gisin99} N. Gisin and S. Wolf, Phys. Rev. Lett., {\bf 83}, 4200 (1999).
\bibitem{Bruss03} D. Bru\ss, M. Christandl, A. Ekert, B. G. Englert,
  D. Kaszlikowski and C. Macchiavello, quant-ph/0303184.
\bibitem{Acin03} A. Ac\'in, N. Gisin and V. Scarani, quant-ph/0303009.
\bibitem{Acin03b} A. Ac\'in, L. Masanes and N. Gisin, quant-ph/0303053.
\bibitem{Sanpera97} A. Sanpera, R. Tarrach and G. Vidal, Phys. Rev. A,
  {\bf 58}, 826 (1997).
\bibitem{Guehne02} O. Guhne, P. Hyllus, D. Bru\ss, A. Ekert,
  M. Lewenstein, C. Macchiavello and A. Sanpera, Phys. Rev. A, {\bf
  66}, 062305 (2002).
\bibitem{Guehne02b} O. Guhne, P. Hyllus, D. Bru\ss, A. Ekert,
  M. Lewenstein, C. Macchiavello and A. Sanpera, quant-ph/0210134.
\bibitem{Bruss98} D. Bru\ss, Phys. Rev. Lett., {\bf 81}, 3018 (1998).
\bibitem{Lewenstein00} M. Lewenstein, B. Kraus, J. I. Cirac and P.
Horodecki, Phys. Rev. A, {\bf 62}, 052310 (2000).
\bibitem{Lewenstein97} M. Lewenstein and A. Sanpera, Phys. Rev. Lett.,
  {\bf 80}, 2261 (1997).
\bibitem{Peres96} A. Peres, Phys. Rev. Lett., {\bf 77}, 1413 (1996).
\bibitem{Horodecki96} M. Horodecki, P. Horodecki and R. Horodecki,
  Phys. Lett. A, {\bf 223}, 1 (1996).
\bibitem{Strang80} G. Strang, {\it Linear Algebra and Its
Applications} (Academic Press, New York, 1980) {\bf }, 2nd ed.
\bibitem{Ekert94} A. Ekert and B. Huttner, J. Mod. Opt., {\bf 41},
  2455 (1994). 
\end{thebibliography}

\end{document}